\documentclass[twocolumn,superscriptaddress,floatfix,prl]{revtex4-2}
\usepackage{graphicx}
\usepackage{dcolumn}
\usepackage{amsmath}
\usepackage{amssymb}
\usepackage{amsbsy}
\usepackage{units}
\usepackage{color}
\usepackage{xcolor}
\usepackage{ulem}
\usepackage{textcomp,txfonts}

\usepackage{pdfpages} 
\usepackage{pgffor} 

\makeatletter
\AtBeginDocument{\let\LS@rot\@undefined}
\makeatother

\def\supplementfilename{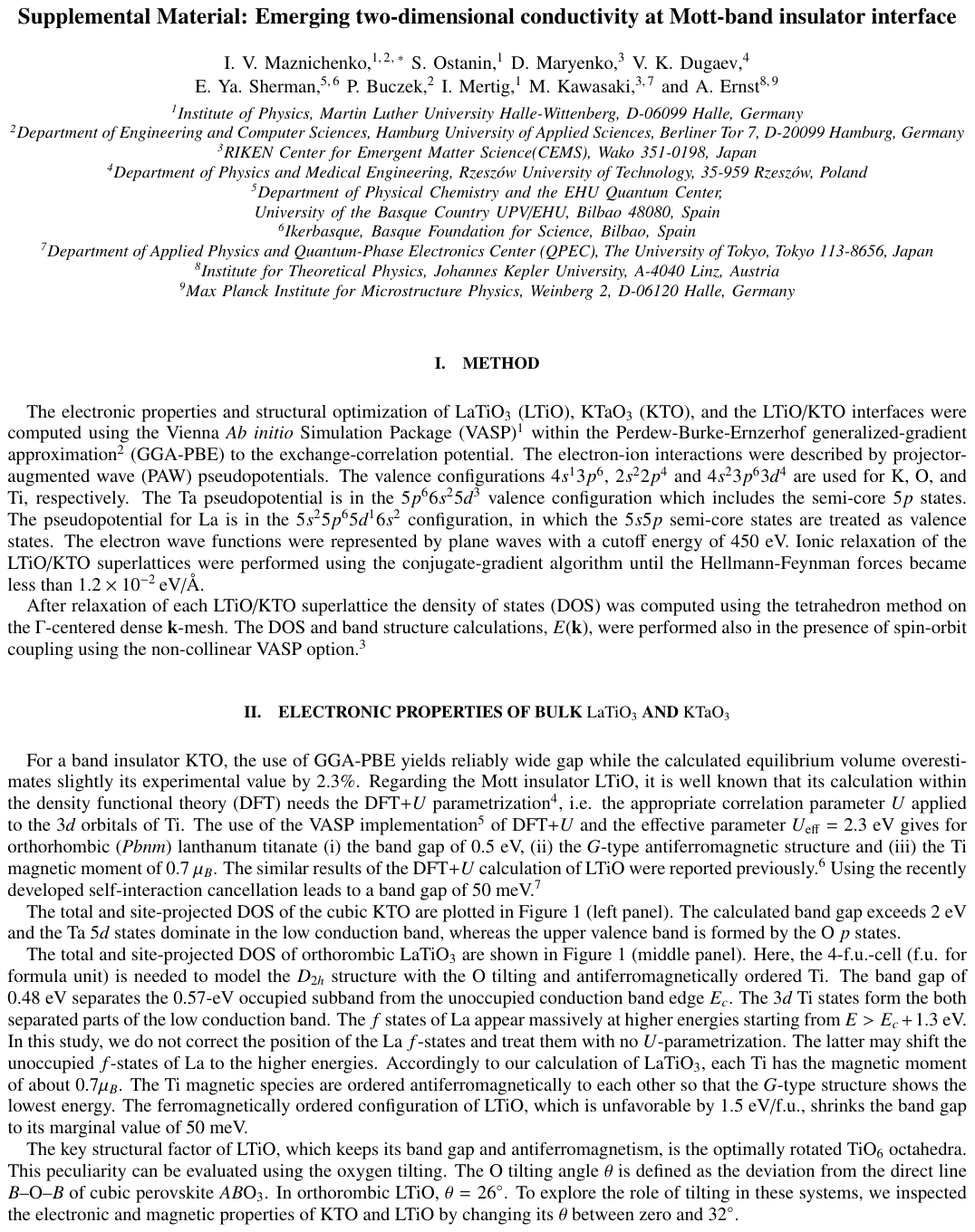}

\pdfximage{\supplementfilename}
\def\numbersupplementpages{\the\pdflastximagepages}

\newif\ifarXiv
\arXivtrue 

\begin{document}

\title{Emerging two-dimensional conductivity at Mott-band insulator interface}

\author{I. V. Maznichenko}
\email{igor.maznichenko@physik.uni-halle.de}
\affiliation{Institute of Physics, Martin Luther University Halle-Wittenberg,
             D-06099 Halle, Germany}
\affiliation{Department of Engineering and Computer Sciences, Hamburg University of Applied 
             Sciences, Berliner Tor 7, D-20099 Hamburg, Germany}

\author{S. Ostanin}
\affiliation{Institute of Physics, Martin Luther University Halle-Wittenberg,
             D-06099 Halle, Germany}

\author{D. Maryenko}
\affiliation{RIKEN Center for Emergent Matter Science (CEMS), Wako 351-0198, Japan}

\author{V. K. Dugaev}
\affiliation{Department of Physics and Medical Engineering, Rzesz\'ow University of Technology, 
 35-959 Rzesz\'ow, Poland}

\author{E.~Ya.~Sherman}
\affiliation{Department of Physical Chemistry and the EHU Quantum Center, University of the Basque Country UPV/EHU, Bilbao 48080, Spain}
\affiliation{Ikerbasque, Basque Foundation for Science, Bilbao, Spain}

\author{P. Buczek}
\affiliation{Department of Engineering and Computer Sciences, Hamburg University of Applied Sciences,
             Berliner Tor 7, D-20099 Hamburg, Germany}

\author{I. Mertig}
\affiliation{Institute of Physics, Martin Luther University Halle-Wittenberg,
             D-06099 Halle, Germany}

\author{M. Kawasaki}
\affiliation{RIKEN Center for Emergent Matter Science (CEMS), Wako 351-0198, Japan}
\affiliation{Department of Applied Physics and Quantum-Phase Electronics Center (QPEC), 
 The University of Tokyo, Tokyo 113-8656, Japan}

\author{A. Ernst}
\affiliation{Institute for Theoretical Physics, Johannes Kepler University, A-4040 Linz, Austria}
\affiliation{Max Planck Institute for Microstructure Physics, Weinberg 2, D-06120 Halle, Germany}

\begin{abstract}

Intriguingly conducting perovskite interfaces between ordinary band insulators are widely explored, 
 whereas similar interfaces with Mott insulators are still not quite understood. Here we address the
  (001), (110), and (111) interfaces between the LaTiO$_{3}$ Mott- and large band gap KTaO$_{3}$ insulators. 
 Based on first-principles calculations, we reveal a mechanism of interfacial conductivity, which 
 is distinct from a formerly studied one applicable to interfaces between polar wide band insulators. 
 Here the key factor causing conductivity is the matching of oxygen octahedra tilting in KTaO$_{3}$ and LaTiO$_{3}$ 
 which, due to a small gap in the LaTiO$_{3}$ results in its sensitivity to the crystal 
 structure, yields metalization of its overlayer and following charge transfer from Ti to Ta. Our findings, also 
 applicable to other Mott insulators interfaces, shed light on the emergence of conductivity observed 
 in LaTiO$_{3}$/KTaO$_{3}$~(110) where the ''polar`` arguments are not applicable
 and on the emergence of superconductivity in these structures.

\end{abstract}

\maketitle

The formation of a conducting layer at the interface between two insulators is one of the most intriguing
problems in the physics of low dimensional electron systems. A remarkable example is the interface between 
two perovskite band insulators LaAlO$_{3}$ (LAO) and SrTiO$_{3}$ (STO)~\cite{Ohtomo2004, Thiel2006}. 
The electrons at such interface demonstrate high mobility enabling the observation of the Shubnikov-de Haas oscillations and the quantum Hall effect~\cite{HaroldLAOSTO, McCollam2014, LAOSTO_QHE}. 
Moreover, such interface can become superconducting, \cite{Reyren2007, LAOSTOSuperconductance2, Li2011,Bert2011}  possibly demonstrating unconventional Cooper pairing 
produced by the Rashba-like spin-orbit coupling.\cite{CavigliaSOC, DaganSOC, BychkovRashba,VenderbosPRX2018, GorkovRashbaSC, UnconventionalSC1,UnconventionalSC2,UnconventionalSC3}  
The understanding, control, and prediction of the emergent phenomena including  
topological properties, quantum geometries, and superconductivity is vitally related to the mechanism of the conducting 
layer formation~\cite{Li2011, Bert2011, Joshua2013, VazNatMat2019,  Lesne2023, QuantumMetrics1,QuantumMetrics2}. 

For oxide electronics, the most prominent mechanism of the conductivity is the so called ''polar catastrophe``,
broadly applicable to the (001)-oriented structures \cite{Nakagawa2006, Hotta2007, Zou, LTOSTONatComm, GTOSTO,LCOKTO,Goniakowski2008, Chen2010a, Bristowe2014, Stemmer2014, Maznichenko2018impact,Maznichenko-PRMat_3_074006_2019,Maznichenko2020formation}. 
Here,  LaAlO$_{3}$  is essentially thought to consist of oppositely charged 
layers [LaO]$^{+}$ and [AlO$_{2}$]$^{-}$, while the respective layers of SrTiO$_{3}$ are electrically neutral. 
The formation of a polar heterojunction AlO$_{2}$/LaO/TiO$_{2}$ eventually results in the electron transfer through the 
LaAlO$_{3}$/SrTiO$_{3}$ interface which causes formation and filling of a conduction band formed by the Ti 3$d$ electron states,
responsible for conductivity and, at appropriate conditions, for superconductivity. 

Recently, the isostructural perovskite KTaO$_{3}$ (KTO) attracted a lot of attention due to its 
ability to produce the surface superconducting state, strongly dependent on the crystal orientation~\cite{KTOUeno, EuOKTOScience, LAOKTOScience, LAOKTO, EuOKTONatComm, EuOKTO110,KTOScienceAdvances, Maryenko-APLMat2023}. Band insulator KTaO$_{3}$
can be brought in contact with Mott insulator LaTiO$_{3}$ (LTiO) epitaxially grown on its surface with emergence of a conducting LaTiO/KTO
interface.\cite{Maryenko-APLMat2023} Seemingly, it would be straightforward to consider 
the ''polar catastrophe`` as the mechanism of the formation of the two-dimensional electron gas (2DEG) at the
interface and the earlier studies for (001)-oriented LaTiO/KTO found consistency with the mechanism.~\cite{Zou} 
This approach, however, fails to explain a highly conducting (110)-oriented interface 
which does not demonstrate the polar discontinuity.   
Moreover, the observed critical dependence of the interface  
superconductivity on the LTiO thickness \cite{Maryenko-APLMat2023} implies that the physics of this 
Mott insulator is one of the key factors responsible for the interface 
electronic properties. Thus, another general mechanism 
of two-dimensional conductivity,
possibly applicable to a variety of band-Mott insulator interfaces is needed.

Here we use first-principles calculations to study the formation of interfacial conducting 
layers for (001), (110), and (111) - oriented LTiO/KTO heterojunctions.  We demonstrate that the crystal 
structure effect, that is the matching in the orientation of the oxygen octahedra surrounding Ti and Ta ions, 
at the interface and the following charge transfer across the interface are 
decisive for the formation of conducting layer. This mechanism resulting from a small gap in the LTiO,
making it highly sensitive to the crystal structure variations,  is distinct from the 
conventional explanation for polar band insulators and can be extended to other 
interfaces with Mott insulators. 

To have reference points, we present the first-principles calculations of the bulk 
materials obtained with the Vienna {\it Ab initio} Simulation
Package (VASP) \cite{Kresse1996} within the Perdew-Burke-Ernzerhof
generalized-gradient approximation \cite{GGA-PBE} (GGA-PBE) for the
exchange-correlation potential and implementation of the lattice relaxation. 
(For technical details see Supplemental Material (SM) \cite{supp}.) For a band insulator KTO, 
having the cubic unit cell with the experimental lattice constant $a_{\rm K}= 3.989$~\text{\AA} \cite{KTOReference},
this approach yields reliably wide gap while the calculated equilibrium 
volume overestimates slightly its experimental value by $2.3 \% $.  
The calculated band gap between Ta 5$d$ and O $p$ states exceeds 2~eV.

Calculation of the  bulk Mott insulator LTiO 
within the density functional theory (DFT) {needs the DFT+$U$ parametrization} \cite{DFT+U,varignon2019origin}, i.e. the
appropriate correlation parameter $U$ applied to the 3$d$ orbitals of Ti.  
This orthorhombic material with lattice parameters $a_{\rm L}=b_{\rm L}=5.595$~\text{\AA}  
and $c_{\rm L}=7.912$~\text{\AA} has a pseudocubic structure 
with the lattice constant $a_{\rm pc}=a_{\rm L}/\sqrt{2} = c_{\rm L}/2= 3.956$~\text{\AA}, different by only 
0.8$\%$ from the $a_{\rm K}$ of the cubic KTO.
This structure is characterized by the tilting angle $\theta=180^{\circ}-\angle B$O$B$ 
(here $B$=Ti), defined as the
deviation from  the cubic perovskite structure $AB$O$_{3}$, with the experimental value $\theta = 26^{\circ}.$ 
The VASP implementation \cite{VASP+U} with the effective 
$U_{\rm eff} = 2.3$~eV gives for orthorhombic ({\it Pbnm}) LTiO: (i) the band gap of 0.5~eV, (ii) the
$G$-type antiferromagnetic structure, and (iii) the Ti magnetic moment
of 0.7 $\mu_{B},$ in agreement with other numerical approaches.\cite{Korean2006,Varignon2019} 
This small 0.5~eV band gap makes the LTiO very sensitive to perturbations 
and, as we will see, can result in the formation of the 2DEG at the LTiO/KTO interface. 

Considering epitaxial interfaces of materials with different lattices, one expects 
a local modification increasing their similarity, within several atomic 
layers near the interface. {Various structural alterations in similar structures were discussed in experimental and theoretical studies  \cite{Ishida2008,okamoto2004electronic,maurice2006electronic,Wong2010,Okamoto2006,
Schoofs_2013,Fister2014}, but the tilting of oxygen octahedra has never been considered as the main source for 2DEG formation.} {However, for small gap materials such as
LTiO, this local modification 
can lead to a sufficient alteration of the electron bands.} Since the principal difference between KTO and LTiO is the 
 tilting of the oxygen octahedra, we expect a certain matching of the octahedra tiltings near the interfaces. 
To clearly show the sensitivity to perturbations in terms of the octahedra tilting, 
we begin with a computer experiment by calculating the hypothetical cubic LaTiO$_{3}$
without TiO$_6$ tilting.   
As this realization is metallic, as shown in the Supplemental Material, 
we see that the key {\it lattice structural} factor of strongly correlated LTiO,
keeping its finite small band gap and antiferromagnetism, is the tilting.


{To quantitatively understand the critical impact of the tilting on the electronic structure and 2DEG formation, we calculate the bulk electronic and magnetic properties of KTO and LTiO at discrete tilting angles without the following relaxation.}

\begin{figure}
   \centering
   \includegraphics[width = 1\columnwidth]{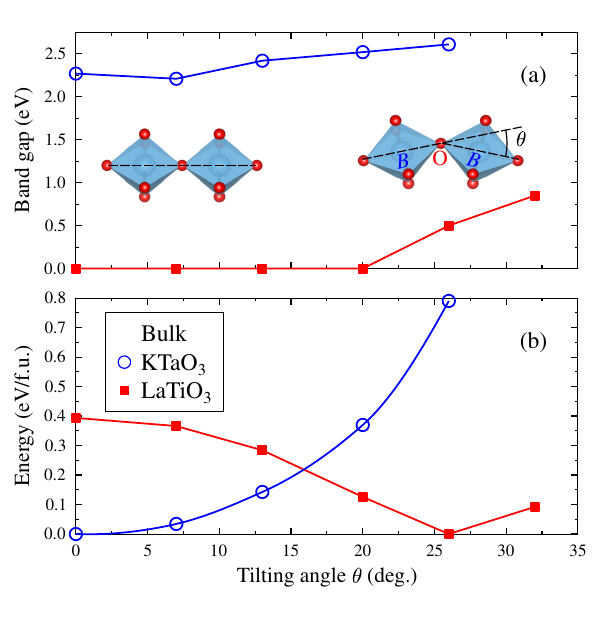}
   \caption{The band gap (a) and total energy (b) of orthorhombic KTO and LTiO as a 
   function of the tilting angle $\theta.$ Inset in panel (a) provides definition of the tilting 
    angle $\theta=180^{\circ}-\angle B$O$B,$ where $B=$Ta and $B=$Ti  for KTO and LTiO, respectively.
    Energies are calculated with respect to the energies with the equilibrium optimally 
    tilted $\theta=0$ for KTO and $\theta= 26^{\circ}$ for LTiO.}
  \label{tilting}
\end{figure}

Figure~\ref{tilting} shows the band gap of KTO and LTiO (panel (a)) and its total energy (panel (b))
calculated as a function of $\theta.$ With decreasing $\theta$ the
LTiO Mott gap decreases and closes for $\theta < 20^{\circ}.$ Thus, a
small decrease in $\theta$ with respect to its experimental value 
in LTiO provokes the insulator-to-metal transition. 
{For a wide-band insulator such as KTO or LAO}, for comparison, the variation in $\theta$ does not affect significantly
the wide gap value.  The total energy of LTiO (Fig.~\ref{tilting}(b))
shows that robustly metallic LTiO calculated without tilting ($\theta=0$) 
is unfavorable by 0.4~eV per the formula unit as compared to the optimally tilted LTiO.  
We found also that the Ti magnetic moments decrease gradually with 
decreasing $\theta$ to completely nonmagnetic Ti sites in untilted LTiO.
The spin-polarized density of states (DOS) of untilted bulk LTiO is presented in the SM \cite{supp}
illustrating that this phase is metallic with 
$E_{F}$ being about 0.8~eV above the conduction band
edge. Thus, if, for certain reasons, the tilting angle in  LTiO at the interface decreases compared to 
the bulk structure, this can cause the 2DEG formation and can be referred to as the ''undertilting`` mechanism 
of emergence of the conductivity. 



\begin{figure}
  \centering
  \includegraphics[width = 1.\columnwidth]{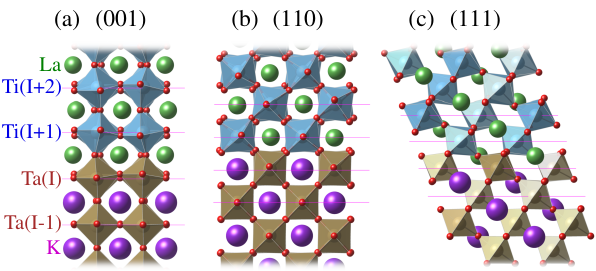}
  \caption{Interfacial LTiO/KTO configurations of the (001), (110), and (111) heterojunctions,
    which are plotted in the (a), (b), and (c) panels, respectively.
    La is shown in green, K in magenta, O in red. The TaO$_6$ and TiO$_6$ octahedra are shown in brown and blue, respectively. 
    A tilting of octahedra is clear to see for TiO$_6$ in LTiO and for interfacial TaO$_6$ in KTO.
    The distances between planes are: $a_{\rm pc}/2 \approx 2$~\text{\AA},  $a_{\rm pc}/(2\sqrt{2}) \approx 1.4$~\text{\AA}, and
$a_{\rm pc}/(2\sqrt{3}) \approx 1.2$~\text{\AA} for (001), (110), and (111), respectively.}
  \label{3-interfaces}
\end{figure}

{We begin with a formal description of the interfaces based on ionic charges.}
Figure ~\ref{3-interfaces} presents 
the interfacial region of the calculated (001), (110), and (111) 
LTiO/KTO structures, respectively. 
{Among three interfaces, only the (110) one is unpolar (see Table~\ref{tab}).}

\begin{table}
  \caption{Plane charge sequence and in-plane magnetic order for three interface orientations. 
    Numbers show the nominal charge for planes ("//" denotes the interface). Note that  the sequence is continuous 
    only for the (110) interface meaning that the polarity-based arguments are not applicable there. 
    The magnetic ordering corresponds to cross-sections
    of the bulk LTiO by the planes with the corresponding orientations.}
\begin{center}
\begin{tabular}{|c|c|l|}
\hline
interface & plane charge sequence & in-plane magnetic order\\
\hline
(001)     & -1/+1/-1/+1//+1/-1/+1/-1    &  AFM chessboard\\
\hline
(110)     & -4/+4/-4/+4//-4/+4/-4/+4    &  AFM chains\\
\hline
(111)     & -5/+5/-5/+5//-3/+3/-3/+3    &  FM\\
\hline
\end{tabular}
\end{center}
\label{tab}
\end{table}

\begin{figure}
  \centering
  \includegraphics[width = 1\columnwidth]{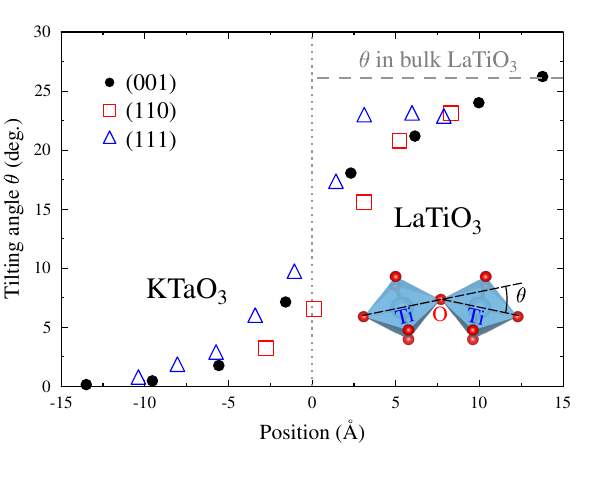}
  \caption{Variation of the tilting angles in LTiO/KTO~(001), (110), and (111)
    interfaces after relaxation. Vertical dashed line accords to $z$
    coordinate of the interfacial atom Ta (I). Inset corresponds to Fig. \ref{tilting}(a).}
  \label{KTO-LTO-tilting}
\end{figure}

{In our main task we study 
the LTiO/KTO interfaces with first principle calculations.} The key feature of the obtained optimized atomic positions and the crystalline structure at the interfaces
is the strong layer-dependence of the tilting angles ${\theta}$ on both LTiO and KTO
sides for all considered orientations. Figure~\ref{KTO-LTO-tilting} shows the calculated
tilting angles ${\theta}$ plotted as a function of the supercell
$z$-coordinate. Although the $z-$dependence of $\theta$ varies from orientation to orientation, as 
seen in Fig.~\ref{KTO-LTO-tilting}, the calculations demonstrate that ${\theta}$ strongly 
decreases in LTiO toward the interface for all of them. The minimization of the lattice energy tends 
to match the orientation of the oxygen octahedra in KTO and LTiO at the interface producing 
KTO-related tilting  up to $10^{\circ}$ in its interfacial unit cells. Thus, the weak tilting of the 
oxygen octahedra in LTiO interfacial layers, insufficient to form the Mott insulator,
leads to its metalization and formation of the 2DEG at all these interfaces. 

\begin{figure*}
  \centering
  \includegraphics[width = 1\textwidth]{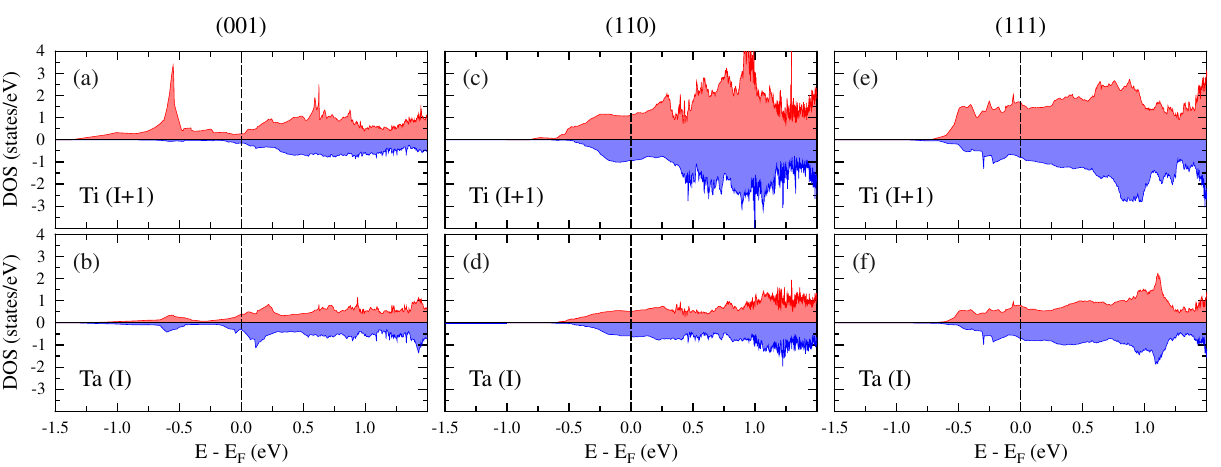}
  \caption{Spin polarized DOS calculated for the interfacial Ti and Ta in
    LTiO/KTO interfaces: (001) -- (a) and (b), (110) -- (c) and (d),
    (111) -- (e) and (f), respectively.}
  \label{KTO-LTO-DOS}
\end{figure*}

Now we consider this interface-based 2DEG in detail. For all considered orientations of interfaces, we
found that it is formed mainly by the $B$-type cations, Ta and Ti,
placed in the two interfacial unit cells of KTO and LTiO with the spin-polarized DOS
calculated for Ta and Ti in the interfacial layers I and I+1 presented in Fig.~\ref{KTO-LTO-DOS}. 
To compare the $3d$ Ti and $5d$ Ta contributions for the three 2DEGs, we calculated
 for Ta(I) and Ti(I+1) the integrated DOS from the conduction band bottom to the Fermi energy $E_{F}.$ 
 The corresponding charges $q,$  which are presented in Table~\ref{tab1}, show that in  
 LTiO/KTO~(001) each interfacial Ta--Ti pair contributes exactly one electron to its 2DEG. 
 Thus, the interface (001) simply closes the gap and transfers the $q$ portion of 1/3 from Ti(I+1) to Ta(I).  
 In LTiO/KTO~(110) and LTiO/KTO~(111), $q$(Ta(I))+$q$(Ti(I+1)) increases to 1.24 and 1.75, respectively.
 This can be attributed to the increased density of states and to the changes in the number of atomic 
 neighbors, with Ta(I) having one, two, and three Ti(I+1) neighbors at the (001), (110), and (111) interface,
 respectively. For all the interfaces, occupations of Ta and Ti sites 
 are mutually related due to the common Fermi energy for all electron subbands and the state 
 hybridization. It is interesting to mention that the calculations show 
 the charge ratio  $q({\rm Ta(I)})/q({\rm Ti(I+1)}) \approx 0.5$ for all of the interfaces.
   
\begin{table}
\caption{Effective charges $q$ (in the units of electron charge) corresponding 
to the 2DEG at LTiO/KTO interfaces, projected at the interfacial Ta and Ti, and corresponding density of states at Fermi level $n_{F}.$}
\begin{center}
\begin{tabular}{|c|c|c|c|c|}
\hline
interface &\ $q$(Ta(I)) \ &\ $q$(Ti(I+1)) \ &  $n_{F}$(Ta(I)) & $n_{F}$(Ti(I+1))  \\
\hline
(001)     & 0.32    &  0.67   & 0.68  &  0.43     \\
\hline
(110)     & 0.39    &  0.85   & 1.10  &  2.03     \\
\hline
(111)     & 0.58    &  1.17   & 1.52  &  2.40      \\
\hline
\end{tabular}
\end{center}
\label{tab1}
\end{table}

The DOS at the Fermi level of the order of 1 state/eV and charges $q({\rm Ta(I)})+q({\rm Ti(I+1)})$ 
correspond to a typical metal with the density of the order of one electron per 
unit cell, in agreement with the experiment.\cite{Maryenko-APLMat2023,Zou}  
For the (110) and (111) interfaces the DOS at the Fermi level is significantly larger than that at the (001)
interface, see  Table~\ref{tab1}. This difference, which for the (110) interface can be attributed to the anisotropy
of the Fermi surface, agrees with the absence of superconductivity in the (001) heterostructures 
\cite{Maryenko-APLMat2023} since the DOS at this interface could be not sufficiently 
large to produce the superconductivity. This argument is generic since a low Fermi level DOS usually 
disfavors formation of Cooper pairs.  

The lower density of states can be attributed to larger interlayer distances in LTiO/KTO~(001)
heterostructure. This reduces hybridization via the interface, which
is reflected in the DOS: the ferromagnetic highly spin-polarized peak at $\approx 0.6$~eV below the Fermi level (see Fig.~\ref{KTO-LTO-DOS}(a,b))
corresponding to narrow bands of correlated Ti-based electrons, includes many electron states.
Since it pulls electrons below the Fermi level and, thus, decreases the corresponding density of states at the Fermi 
level as expected for the (001)-related symmetry. 
In addition to the charge transfer, the proximity with the highly polarized Ti(I+1) magnetic moments 
induces a weak spin polarization of neighboring Ta with the magnetic 
moments of the interfacial Ta(I) being about 0.06~$\mu_{B}$ {(see Sec. III in SM for analysis of atomic configurations at interfaces \cite{supp}).} 

\begin{figure*}
  \centering
  \includegraphics[width = 1.0\textwidth]{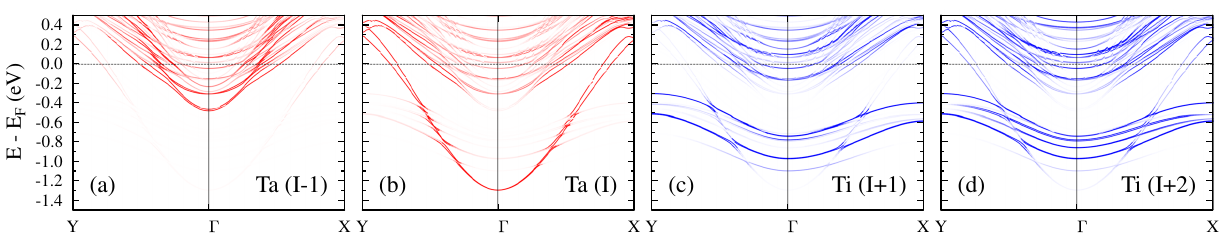}
  \caption{The 2DEG band structure associated with interfacial Ta and Ti in LTiO/KTO~(001)
  with the notations defined in Fig. \ref{3-interfaces}. The large density of Ti-based bands at $E-E_{F}\approx -0.6$~eV 
  corresponds to the peak in the density of states in this energy range as presented in Fig. \ref{KTO-LTO-DOS}. The intensity of the lines
  corresponds to the contribution of the given atom in the band state. }
  \label{KTO-LTO-001bands}
\end{figure*}

The very similar shapes of their site-projected DOS and similar numerical values, 
show that the metalization involves Ti and Ta 
atoms contributions on the same scale, with $5d$ of Ta states getting
a considerable population and becoming conducting together with the contribution
of Ti electrons. Thus, the origin of the 2DEG in
these systems is different from the case of the LAO/STO~(001)
interface, in which the 2DEG is formed with the interface 
polarity accompanied by corresponding band bending. 

In  Fig.~\ref{KTO-LTO-001bands} each panel shows the energy branches
associated with the $d$ states of Ta or Ti near the interface.
The lowest $E(\bf{k})$ branches of LTiO/KTO~(001) which form its
conduction band edge, $E_{C}$, belong to the interfacial Ta(I), as
Fig.~\ref{KTO-LTO-001bands}(b) shows.  Figures \ref{KTO-LTO-001bands}(a) and \ref{KTO-LTO-001bands}(b) 
show the up-shift of the bottom of the corresponding conduction band from Ta(I) to Ta(I-1) layer 
by approximately 1~eV, indicating a very strong band bending effect, common with the 
LAO/STO~(001). For comparison, the branches of interfacial Ti(I+1) and the next to it Ti(I+2) 
appear at approximately the same energies, meaning a much weaker band bending effect on this side of the interface, 
as can be seen in Figs. \ref{KTO-LTO-001bands}(c) and \ref{KTO-LTO-001bands}(d). 
No significant band bending was numerically found at the LTiO/KTO~(110) and (111) interfaces. 
Comparison of Figs. \ref{KTO-LTO-001bands}(a)-\ref{KTO-LTO-001bands}(d) shows that 
2DEG  at the LTiO/KTO~(001) interface is formed by three different kinds of electron bands:
(i) broad Ta-based bands weakly hybridized with the Ti-based ones, demonstrating 
a considerable band banding and contributing to the formation of the 2DEG, 
(ii) a large number of narrow Ti-based bands, not crossing the Fermi level,
weakly hybridized with the Ta-based ones, and (iii) a large number of considerably 
hybridized Ti and Ta- based bands crossing the Fermi 
level. The two latter kinds of bands do not show a strong bending pattern.

Summarizing, the (001), (110), and (111) crystal interfaces between
the polar materials such as Mott insulator LaTiO$_{3}$ and wide band KTaO$_{3}$ were simulated from
the first principles.  For all three interfaces we
found that their calculated metallic densities of states, formed mostly by the interfacial Ti 3$d$  and Ta 5$d$ states,
qualitatively agree with the experimental results. One of key reasons for the
formation of two-dimensional metals in all these systems is a strong altering
of the oxygen octahedra  tilting angles at the interfaces, matching their orientation 
in KTO and LTiO and considerably decreasing it at the LTiO side compared to the corresponding
bulk value. This ''undertilting`` destroys the small LTiO  Mott-like 
band gap {at all interfaces (see Figs. \ref{tilting} and \ref{KTO-LTO-tilting})}, making it the qualitative feature for {these systems}. At the (001) and (111) interfaces
this mechanism {can work together with} the polarity-induced interface charge transfer making these two effects
involving interacting electrons and lattice distortion, inseparable. 
However, it is important to stress that the appearance of the conducting electrons at the (110) 
interface cannot be attributed to the polarity (see Table \ref{tab}) effects and, 
therefore, the ''undertilting``  is critically important for the conductivity and, 
at appropriate conditions, for the superconductivity, of this heterostructure. 
We note that the relatively high density of states  of the conducting electrons at the (110) and
(111) interfaces can be the decisive factor of their superconductivity in contrast 
to the (001) interface which is metallic but not superconducting. Another factor which can be detrimental 
for superconductivity at the (001) interface is the ferromagnetic behavior of interfacial electron 
states shown in Fig. \ref{KTO-LTO-DOS}(a).
{The role of local defects in the 2DEG formation
was not considered in our calculations since for the obtained 2DEG at
the entire interface area with a large electron concentration and the
density of states, a weak disorder in high-quality structures is a
marginal effect. \cite{Maznichenko2018impact,Maznichenko-PRMat_3_074006_2019,Maznichenko2020formation,Maryenko-APLMat2023}.}

To provide an outlook and further development of this research, we stress that
the proposed picture of formation of two-dimensional conducting systems of interacting electrons
can be applied to various interfaces between wide gap- and small gap- perovskite 
Mott insulators. To make this picture applicable, the interface structure 
should favor tilting of the oxygen octahedra different 
from that in the bulk Mott insulator with its band structure being 
strongly sensitive to the tilting, as possibly can occur for thin films 
of LaVO$_{3}$ grown on SrTiO$_{3}$ substrates. \cite{Hotta2007,Rotella2012}


\begin{acknowledgments}
A.E. acknowledges funding by Fonds zur F\"{o}rderung der Wissenschaftlichen Forschung (FWF) Grant No. I 5384.
The work of E.S. is financially supported through Grant No. PID2021-126273NB-I00 
funded by MCIN/AEI/10.13039/501100011033 and 
by ERDF “A way of making Europe,” and by the Basque Government through 
Grant No. IT1470-22.
\end{acknowledgments}

\bibliography{LTO-KTO_paper}

\end{document}